\documentclass[fleqn,usenatbib]{mnras}
\RequirePackage{fix-cm}
\usepackage{mathptmx}
\usepackage{times,txfonts}

\usepackage[T1]{fontenc}
\usepackage{ae,aecompl}

\usepackage{graphicx}	
\usepackage{comment}

\newcommand{\mjyb}{mJy beam$^{-1}$\,}

\title[Fourth arc in Abell 2626]{Discovery of a fourth arc in Abell 2626 at 610 MHz with the GMRT: Spectral properties and possibilities for the origin}

\author[Kale R.]{Ruta Kale,$^{1}$\thanks{ruta@ncra.tifr.res.in} and Myriam Gitti$^{2,3}$
\\
$^{1}$National Centre for Radio Astrophysics, Tata Institute of Fundamental Research, Post Bag 3, Pune 411007, India\\
$^{2}$ Physics and Astronomy Department, University of Bologna, via
Ranzani 1, 40127 Bologna, Italy \\
$^{3}$INAF, Istituto di Radioastronomia di Bologna, via Gobetti 101, I-40129
Bologna, Italy \\
}

\date{Accepted XXX. Received YYY; in original form ZZZ}

\pubyear{2016}

\begin{document}
\label{firstpage}
\pagerange{\pageref{firstpage}--\pageref{lastpage}}
\maketitle

\begin{abstract}
We report the discovery of a fourth eastern arc (Arc E) towards the cool-core cluster Abell 2626 using 
610 MHz Giant Metrewave Radio Telescope observations. 
Three arcs towards north, west and south were known from earlier works at 1400 MHz and 
proposed to have originated in precessing radio jets of the central active galactic nucleus. 
The 610 - 1400 MHz integrated spectral indices of the arcs are in the range 3.2 - 3.6 and the spectral index 
map shows uniform distribution along the lengths of the arcs. 
If associated with A2626, the arcs have linear extents in the range 79 - 152 kpc. 
The detection of Arc E favours the scenario in which a pair of bipolar precessing jets 
were active and halted to produce the arc system. Based on the morphological symmetry and spectral similarity, 
we indicate a possible role of gravitational lensing. Further high resolution low frequency observations and measurements 
of the mass of the system are needed to disentangle the mystery of this source.
\end{abstract}

\begin{keywords}
gravitational lensing:strong -- radiation mechanisms:non-thermal -- galaxies:clusters:individual:Abell 2626 -- radio continuum:general  
\end{keywords}



\section{Introduction}\label{intro}
 Radio sources at the centres of 
 clusters of galaxies show complex morphologies involving variation of intrinsic properties 
 and modification by interactions with environments. Cavities in the intra-cluster medium carved 
 by radio lobes \citep{mcn07,2012NJPh...14e5023M,2012AdAst2012E...6G,2012ARA&A..50..455F} 
 and the binary radio galaxies, as the spectacular ones observed in A400 \citep[3C75,][]{1985ApJ...294L..85O} 
 are examples of such phenomena. A recent discovery of a complex source is the ``Zwicky's Nonet'' which are precessing radio jets in a dense 
 environment \citep{biju16}. 
 
 In this work we focus on the puzzling radio arcs in the cluster Abell 2626  (hereafter, A2626).
 The galaxy cluster A2626 is at a redshift of 
0.0553 \citep{1999ApJS..125...35S} and known to be a cool-core cluster 
\citep{1997MNRAS.292..419W}.
It has an X-ray luminosity of $1.44\times10^{44}$ erg s$^{-1}$ \citep{2000ApJS..129..435B}. 
The brightest cluster galaxy (BCG) at its centre is a cD galaxy (IC5338) with 
two nuclei of which the south-west (SW) nucleus shows emission lines \citep{Crawford99}.

This cluster was known to have a central radio source with surrounding 
diffuse emission of steep ($\alpha \sim 2.2$) 
radio spectrum\footnote{The spectral index, 
$\alpha$ is defined as $S_\nu \propto \nu^{-\alpha}$, where 
$S_\nu$ is the flux density at the frequency $\nu$.} \citep{1985A&A...148..323R}.
The diamond-shaped radio source noticed in the early observations of 
this cluster was confirmed as two distinct bars and a mini-halo 
around the central BCG by \citet[][hereafter, G04]{git04}.
Their 330 MHz Very Large Array (VLA) observations also confirmed that the bar-like features 
had very steep spectral indices ($2 - 3$). A deep X-ray study of this cluster using 
the {\it XMM Newton} and {\it Chandra} data showed surface brightness drop at the edges of the proposed 
mini-halo \citep{2008ApJ...682..155W}. The bars on the north and south were proposed to be 
remnants of precessing jets from the AGN in the BCG. 
Deeper and higher resolution VLA observations by 
\citet[][hereafter G13]{git13} at 1400 and 4800 MHz revealed that the central compact radio source 
 associated with the SW nucleus of the BCG had jet-like extensions. 
 Furthermore, the radio bars were shown to have a clear arc-like morphology, and an additional arc to the west 
 was also discovered.
The peculiarities of the radio arcs are that 
they are convex towards the cluster centre unlike typical ICM shock-like features or any gravitational lensing 
arcs. Models involving precessing jets from the two supermassive black holes (SMBHs) at the two nuclei of the 
BCG were invoked to explain the source. Overall, the previous studies have not reached a firm conclusion on the origin 
of this source and referred to it as a puzzle.

We present the analysis of Giant Metrewave Radio Telescope (GMRT) data 
at 610 MHz that leads to the discovery of the fourth eastern arc in the system and allows 
spectral index maps of the arcs between 610 - 1400 MHz for the first time.
The TIFR GMRT Sky Survey Alternative Data Release\footnote{\url{http://vo.astron.nl/tgssadr/q_fits/cutout/form}} 153 MHz image 
\citep{2016arXiv160304368I} are also used to study the radio arcs.
The paper is organised as follows: The observations and data analysis are described in Sec.~\ref{obs}. 
The properties of the arcs are presented in Sec.~\ref{radioarcs}. 
The precessing jets model is discussed in Sec.~\ref{prejets} and motivation for an alternative model involving gravitational lensing 
is provided in Sec.~\ref{gl}. The summary and conclusions are presented in Sec.~\ref{sumcon}.

We use $\Lambda$CDM cosmology with $H_0$=$70$ km s$^{-1}$ Mpc$^{-1}$, $\Omega_\Lambda$=$0.73$ and 
$\Omega_m$ = $0.27$. At the redshift 0.0553 of the cluster A2626, 1 arcsec corresponds 
to 1.08 kpc \citep{coscal}.

\begin{center}
 \begin{figure*}
\label{arcs}
\includegraphics[trim=0cm 0.4cm 0cm 1cm,clip,height=6.8cm]{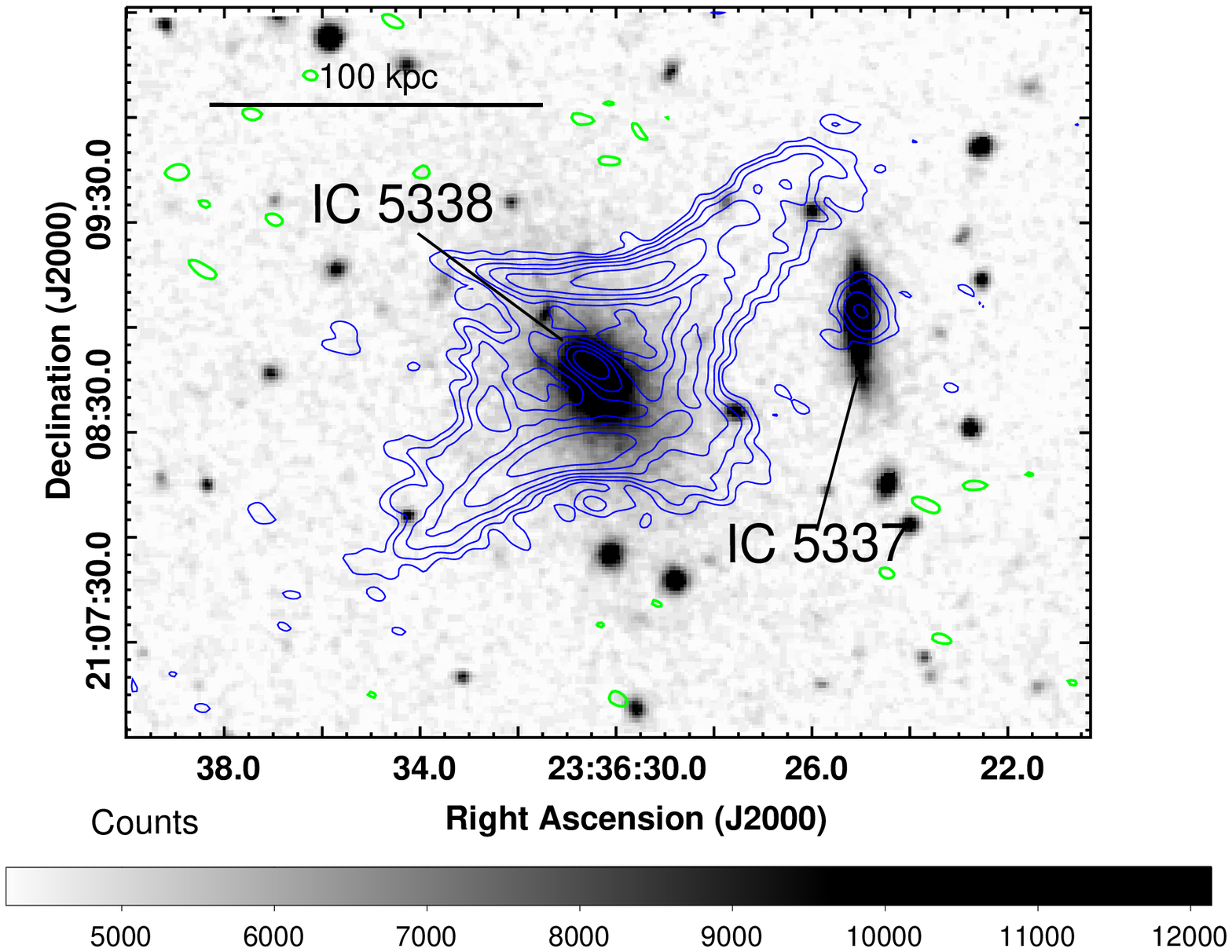}
\hspace*{0.1cm}
\includegraphics[trim=0cm 0.4cm 0cm 1cm,clip,height=6.8cm]{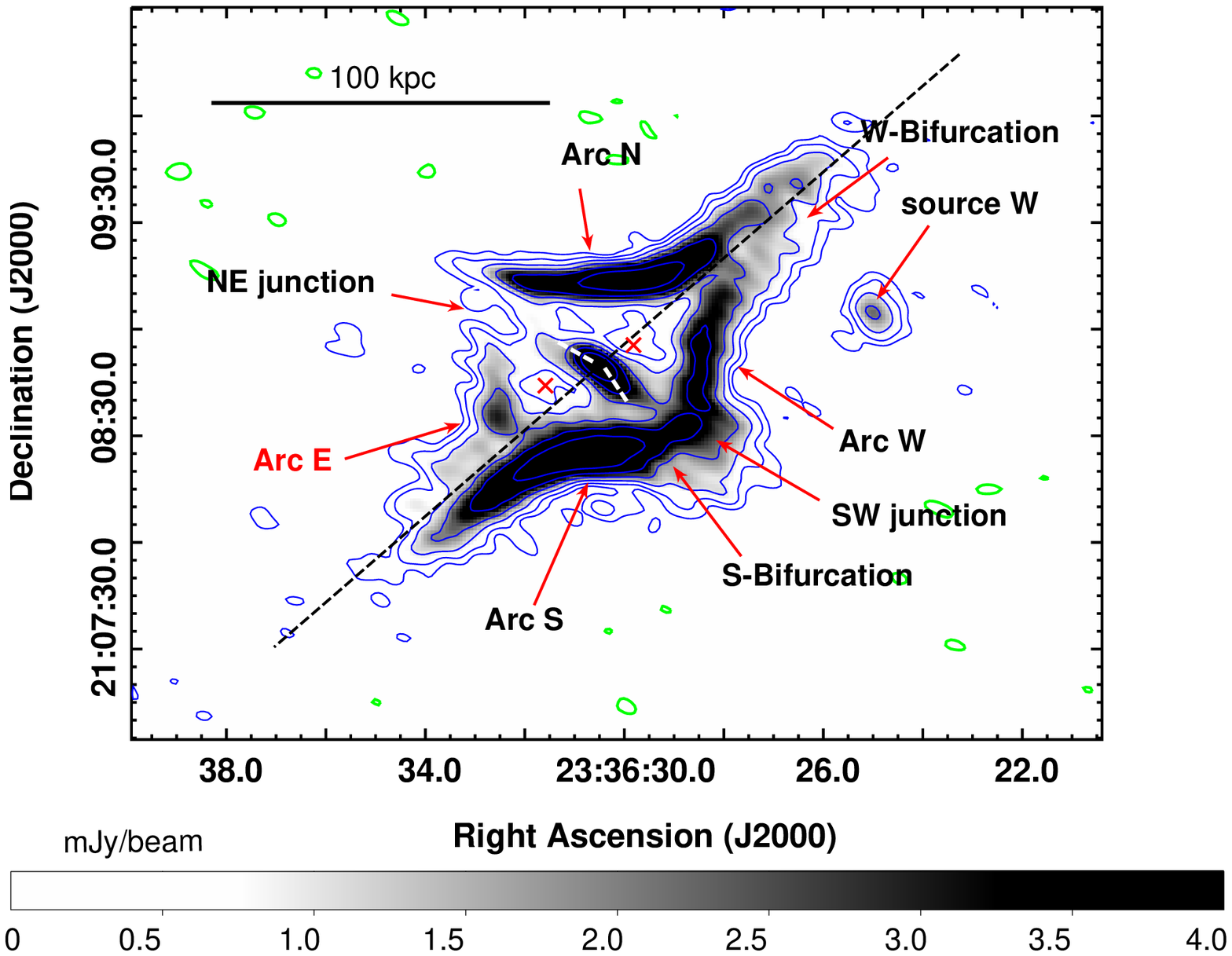}
\caption{ {\it Left --} GMRT 610 MHz image of A2626 shown in contours levels at 
$3\sigma \times [\pm 1,\, 2,\, 4, ...]$ \mjyb where $\sigma = 0.08 $\,\mjyb, overlaid on the 
Digitised Sky Survey R-band image in grey-scale. The blue contours 
are positive and green are negative. The GMRT image has a beam of $8.5^{\prime\prime}\times4.4^{\prime\prime}$ with 
a position angle $61.8^{\circ}$.
{\it Right --} GMRT 610 MHz image shown in greyscale and in contours levels 
as in the left panel. The labels show the A2626 arc system discussed in the text.
The crosses mark two low brightness regions on either side of the central source. The black dashed line 
marks the line where the arcs separate. The white dashed line shows the directions of the jets.
}
\end{figure*}
\end{center}

\section{Observations and data analysis}\label{obs}

\begin{table*}
\begin{center}
\caption[]{\label{obstab} Summary of radio observations.}
\begin{tabular}{cccccccc}
\hline\noalign{\smallskip}
Telescope,&Date & Freq.& Band Width   &Time &Beam&rms \\
Project code         & &MHz  & MHz & min. &$^{\prime\prime}\times^{\prime\prime}$, p. a. &mJy beam$^{-1}$\\
\hline\noalign{\smallskip}
GMRT, 01TCA01    &05-May-2002 &610 & 16 & 90&$8.5\times4.4$, $61.8^{\circ}$ &0.08\\
\hline\noalign{\smallskip}
\end{tabular}
\end{center}
\end{table*}

\begin{figure*}
\label{spix}
\includegraphics[trim=0cm 0cm 0cm 1cm,clip,height=7.2cm]{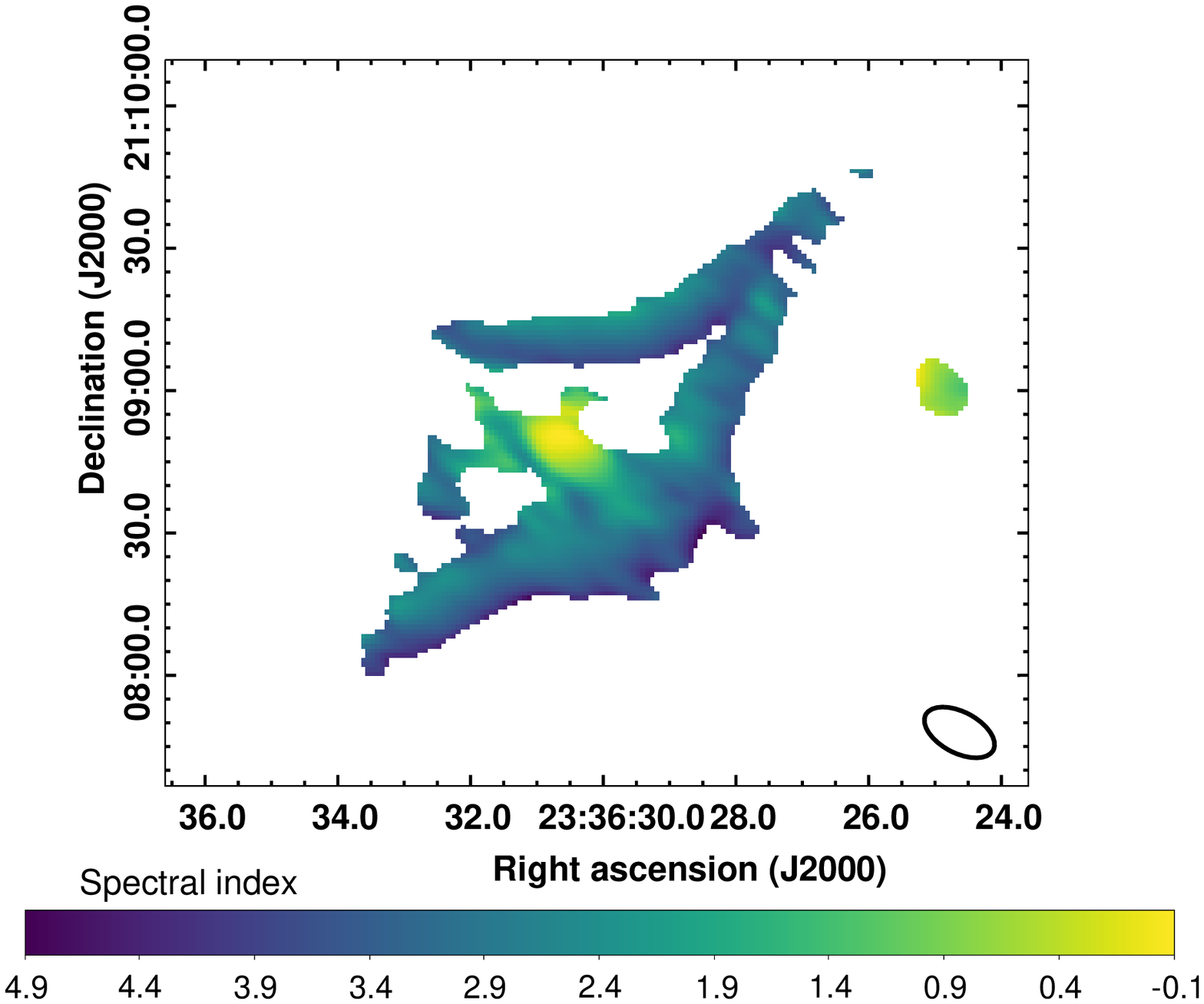}
\hspace*{0.1cm}
\includegraphics[trim=0cm 0cm 0cm 1cm,clip,height=7.2cm]{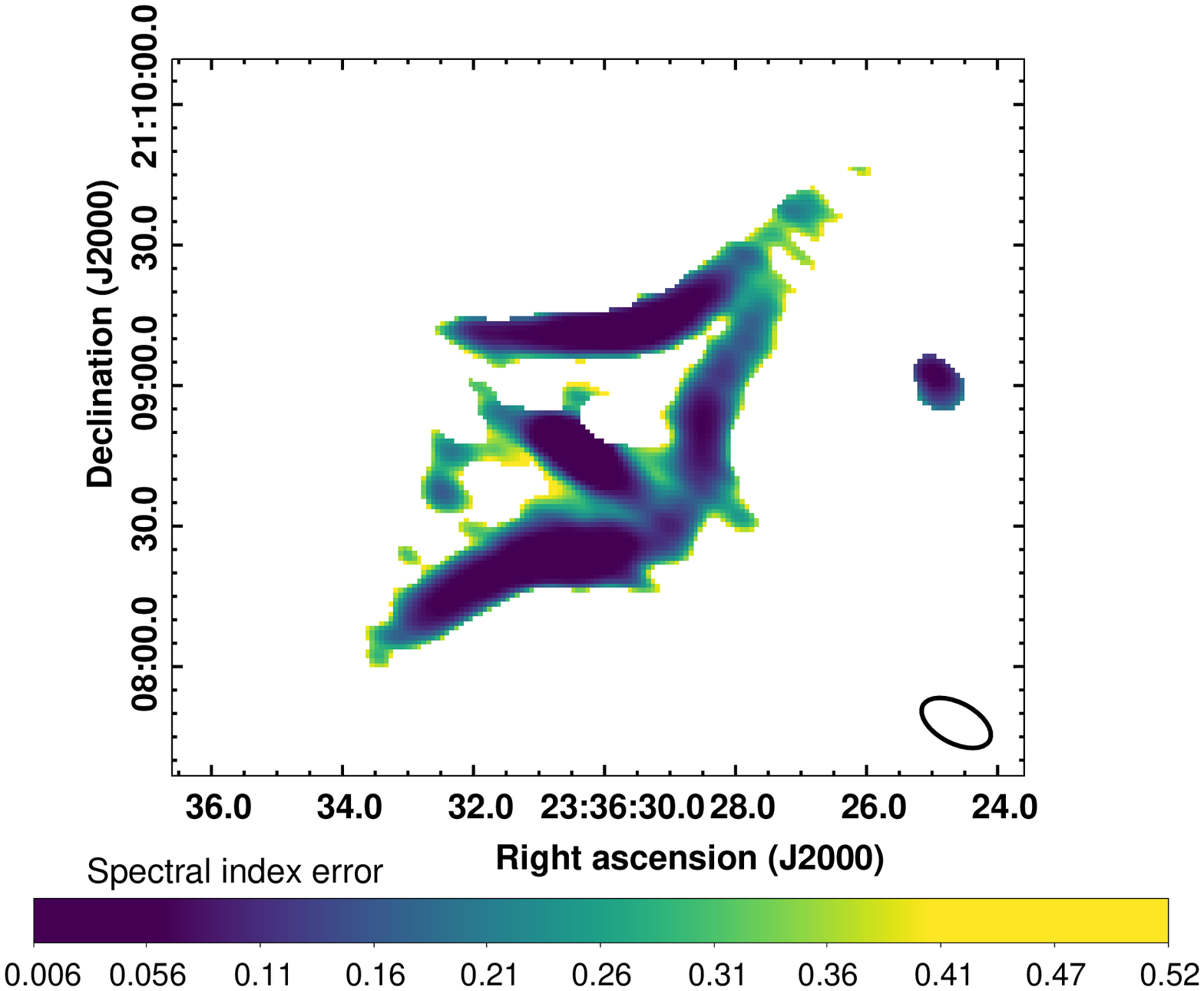}
\caption{Spectral index map between 610 - 1400 MHz ({\it left}) and the corresponding error map ({\it right}) 
are shown.  The maps have a resolution of $8.0^{\prime\prime}\times4.3^{\prime\prime}$ with a position angle $62.4^{\circ}$ shown by the ellipse.}
\end{figure*}

We have used radio observations at 610 MHz from the GMRT 
(see Table ~\ref{obstab} for the observation setup). 
The shortest baseline of 100 m 
at the GMRT enables sampling of the sky at the largest angular scale of $17'$ at 610 MHz 
and is suitable to probe  
 diffuse emission on the scales of a few hundred kpc at the redshift of A2626.
The data were analysed using the NRAO Astronomical Image Processing System (AIPS). 
The standard steps of flagging (excising bad data) and calibration were carried out.
 Absolute flux calibration was carried out using 
the source 3C48. The sources 3C48 and 3C147 were used for bandpass calibration. 
The calibrator 2254+247 was used to calibrate the phases towards the target source.
The calibrated visibilities on the target were imaged and a few rounds of phase-only 
self-calibration were carried out to improve the image sensitivity.
The final image (Fig.~\ref{arcs}) was made using robust 0  weights for the visibilities. 
It has a beam of $8.5^{\prime\prime}\times4.4^{\prime\prime}$ with a position angle $61.8^{\circ}$ and rms noise of 
$0.08 \,$ \mjyb.

Spectral index map between 1400 and 610 MHz was made using the  
VLA A+B array data (G13) and the GMRT 610 MHz data. Images were made using 
the overlapping uv-coverage between 1 - 50 k$\lambda$ at both the frequencies to avoid 
artefacts due to unequal sampling in spatial frequencies. 
The images were corrected for the respective primary beam gains. The 1400 MHz image was 
convolved to match the resolution of the  610 MHz image ($8.0^{\prime\prime}\times4.3^{\prime\prime}$, position angle $62.4^{\circ}$) 
and the resulting images were blanked 
in regions having flux density below $3\sigma$ ($0.06$ and $0.21$ \mjyb
at 1400 and 610 MHz, respectively). 
These images were used in AIPS task COMB to make the spectral 
index map and the corresponding error map (Fig.~\ref{spix}).

\section{Radio arcs in A2626: Discovery of Arc E} \label{radioarcs}
Four distinct arcs are detected at 610 MHz that surround the central core-jet system 
at the BCG, IC 5338 (Fig.~\ref{arcs}). The Arcs N, S and W were previously known (G13) and the 
eastern arc (Arc E) is revealed for the first time. 
The Arc E has an extent of $\sim 79$ kpc along the arc and a flux density of $14.4\pm1.5$ mJy at 610 MHz.
We convolved the 1400 MHz image (VLA A+B array) by G13 to the beam of 610 MHz 
and found a flux density of 0.7 mJy in the region matching that of Arc E at 610 MHz.
The flux densities, extents at 610 MHz and spectral indices between 1400 and 610 MHz 
of the arcs and the core are reported in Tab.~\ref{arcprop}. 
The central core-jet system appears connected with a low brightness bridge along NE-SW axis 
to the arcs.

The 610 - 1400 MHz spectral index map (Fig.~\ref{spix}) shows the flat spectrum core and the steep 
spectrum arcs. The spectra show no rapid changes along the lengths of the arcs. We are limited 
by resolution to interpret any trend along the widths of the arcs. The SW-junction shows a smooth 
steepening trend from the central source into the Arcs W and S.

\begin{figure}
\label{tgss}
\includegraphics[trim=0cm 0cm 0cm 0cm,clip,height=7.2cm]{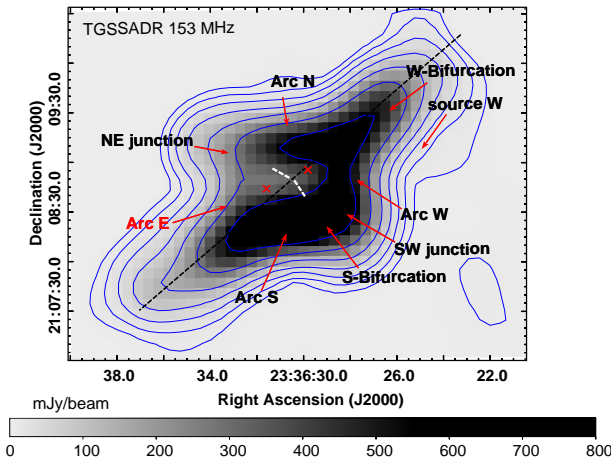}
\caption{The TGSSADR 153 MHz image with a resolution of $25^{\prime\prime}\times25^{\prime\prime}$ and rms noise of 2.8 
mJy beam$^{-1}$ is shown in contours and in greyscale. The contours are at $2.8\times[\pm3, 6, 12, ...]$ \mjyb.
The labels from Fig.~\ref{arcs} are shown for reference.}
\end{figure}

The arcs are also detected in the TGSSADR 153 MHz image with an rms of 2.8 \mjyb and a resolution 
of $25^{\prime\prime}\times25^{\prime\prime}$ (Fig.~\ref{tgss}). The overall diamond shape is detected but the 
resolution is not sufficient to resolve the individual arcs completely. We convolved the 
610 MHz image with a beam of $25^{\prime\prime}$ and compared the flux densities in matched regions on the arcs.
The flux ratios between 153 and 610 MHz are comparable to that between 610 and 1400 MHz (Table ~\ref{arcprop}). 
This shows that the arcs follow the same spectral index as between 610 and 1400 MHz up to 153 MHz. 

\begin{table*}
\centering
\caption[]{\label{arcprop}Properties of the radio sources. The size of the arc is the maximum extent along its length 
at 610 MHz. The flux densities, S$_{153\mathrm{MHz}}$ and S$_{610\mathrm{MHz}}^{'}$ are 
determined from matched regions in the images at resolutions of $25^{\prime\prime}\times25^{\prime\prime}$.\\}
\begin{tabular}{lcccccccc}
\hline\noalign{\smallskip}
Source&S$_{1400\mathrm{MHz}}$&S$_{610\mathrm{MHz}}$& $\alpha_{610}^{1400}$&Size&S$_{153\mathrm{MHz}}$ &S$_{610\mathrm{MHz}}^{'}$ \\
& mJy &mJy&& kpc & mJy & mJy \\
\hline\noalign{\smallskip}
Arc N &$7.0\pm0.4$&$101\pm10$&$3.2\pm0.1$&152 &2200 & 138 \\
Arc S &$9.9\pm0.5$ &$154\pm15$&$3.3\pm0.1$&120&2290 &139\\
Arc W &$3.4\pm0.2$ &$49.6\pm5.0$&$3.2\pm0.1$&97&2230 &111\\
{\bf Arc E} &$0.7\pm0.1$&$14.4\pm1.5$&$3.6\pm0.2$&79&1020&55\\
Core+jets & $19.1\pm1.0$ &$33.1\pm3.3$&$0.7\pm0.1$&29&-&\\
\noalign{\smallskip}
\hline\noalign{\smallskip}
\end{tabular}\\
\end{table*}

The properties of the arcs system detected at 610 MHz are described below.
 \begin{enumerate}
  \item Arc S is the brightest in radio flux density followed by 
 Arcs N, W and E in descending order of brightness at both 610 and 1400 MHz. 
 \item Arc N is the longest ($152$ kpc) and Arc E is the shortest ($79$ kpc) in extent.
 \item The Arcs S and W show broadening towards their junction. The western end of the southern 
 arc near the SW junction is broader and shows evidence of bifurcation. The broadening is also 
 seen in the $1.2^{\prime\prime}$ resolution 1400 MHz image shown in Fig.1 in G13. 
 \item Arc N has the smallest width as compared to the other arcs. It is unresolved along the width 
 at $1.2^{\prime\prime}$ resolution at 1400 MHz (Fig. 1, G13).
 \item The Arcs S and W form the bright SW-junction. The Arcs N and W and Arcs E and S do not appear to 
 join; a distinct low brightness region is detected at 610 MHz between these, shown by the black dashed line in Fig~\ref{arcs}.
 The NE junction is the low brightness region that connects the central source and the Arc N, but the Arc E is not detected 
at the junction.
 \item There are regions of low brightness on the east and west of the central source (crosses in Fig.~\ref{arcs}, right).
 \item The arc system has an extent of about $2^\prime55^{\prime\prime}$ or $189 $ kpc along the NW-SE axis and 
 about $1^\prime35^{\prime\prime}$ or $102$ kpc along the NE-SW axis. 
 \item The axes of the diamond shape formed by the arcs are oriented at an angle of $100^\circ$ with each other.
\item As measured along the N-S direction from the radio core of IC 5338, the Arcs N and S are at angular distances of 
 $24^{\prime\prime}$. Similarly along E-W direction the Arcs E and W are at about $29^{\prime\prime}$ from the radio core.
 \item The spectral indices of the Arcs N, S and W are about 3.2 within $1\sigma$. The steeper spectrum of the 
 arc E may be due to loss of flux density detected at 1400 MHz (VLA A+B array) due to insufficient 
 short baseline coverage. 
 \end{enumerate}

\section{Origin of the radio arcs}

\subsection{Precessing jets ?}\label{prejets}
Past studies have proposed a model involving precessing bipolar jets from the central AGN  
to explain the arcs in A2626 \citep[][G13]{2008ApJ...682..155W}. Such a scenario requires the precession  
axis approximately along the north-south with a halting of jets 
resulting in the steep spectrum ageing remnant arcs (Arcs N and S). 
The explanation of the observed curvature of the Arcs N and S needs further fine tuning of the time of jet halt and 
its position along the precession cone. 
The western arc was proposed by G13 as a possible buoyantly risen bubble or a third remnant of jet activity along 
the east-west. The detection of the Arc E favours the occurrence of another system of bipolar precessing jets with 
precession axis along the east-west. The jet activity producing the N-S and the E-W pairs of arcs may be 
due to the same AGN or involving also the NE nucleus of the BCG.
If the same AGN produces the two activities then a rapid jet reorientation, 
such as plausible in the event of a binary SMBH  
coalescence \citep[][]{2002Sci...297.1310M}, is required. 
If there were a considerable time gap between the start and halt 
of the activity that produced the N-S and the E-W arc pairs, 
 the different ageing should produce steeper spectra in the older arc pair.
The N-S arcs are brighter than the E-W arcs but have comparable spectral indices (Table ~\ref{arcprop}). 
Observations with arcsec scale resolutions at frequencies < 610 MHz are needed to locate any breaks in the 
spectra of the arcs to consolidate the age differences in them if any. 
The NE nucleus does not show emission lines characteristic of AGN \citep{Crawford99} 
and is not active in radio (G13). The role of NE nucleus is not supported by evidence but cannot be 
ruled out with the present data.
\subsection{Gravitational lensing ?}\label{gl}
 Motivated by the remarkable morphological symmetry and spectral similarity of the arcs, we invoke a possibility
 other than the plausible but rare event of precessing double black hole jets.
 When light from a background source is bent due to the gravity of 
a massive object along the line of sight resulting in multiple distorted images of 
the background source, it is termed as gravitational lensing 
\citep[see][for a review]{2010CQGra..27w3001B}. 
Complex arcs and loop-like morphologies 
have been observed and explained as gravitationally lensed radio galaxies \citep[e. g. PKS 1830-211][]{1993ApJ...407...46N}. 
Depending on the geometry of the lens and the source morphology, arcs with curvatures opposite to that in standard lenses 
can occur \citep{2010cosp...38.2299L,2011AstL...37..233L}.
If the arcs were multiple images of a single source, then the flux ratios between 
the arcs are expected to be the magnification ratios \citep[e.g.][]{1992grle.book.....S}; these are in the range 
0.3 - 10.7 for the arcs (Table ~\ref{arcprop}). Higher resolution observations at low frequencies such as 
possible with the LOw Frequency ARray \citep[LOFAR,][]{2013A&A...556A...2V} are needed to further constrain 
the spectral indices in the arcs.
Using the properties of the lens system based on radio images, detailed modelling of the central galaxy as 
a lens and constraints on the background source are needed to consolidate the possibility of gravitational lensing.
\section{Summary and conclusions}\label{sumcon}
We have presented high resolution low frequency observations of the cluster A2626 that is known 
to have puzzling radio arcs. The GMRT 610 MHz observations have led to the discovery of the fourth, eastern arc 
that completes a peculiar diamond shaped system of arcs towards A2626. The arcs have steep 
spectral indices of 3.2 - 3.6. The spectral index map between 610 - 1400 MHz shows uniform 
distribution of spectral indices along the lengths of the arcs; we are limited by resolution along the widths of the arcs.
The arcs could be caused by a rare phenomenon involving a pair of bipolar radio jets precessing about nearly perpendicular 
axes and halting. However the sharp morphology of 
the arcs unlike diffuse remnant lobes of radio galaxies, the curvature of the arcs and little signature of spectral 
variation in the arcs need to be explained in detail.
We indicate a bizarre possibility of the arcs being gravitationally lensed 
images of a background extended source drawing attention to the striking morphological symmetry and spectral similarity in the arcs.
To validate the proposed scenarios, detailed modeling using the properties presented here and derived from higher resolution  
observations are needed.
\section*{Acknowledgements}
We thank the referee and the editor for their comments.
We thank the staff of the GMRT that made these observations possible. 
GMRT is run by the National Centre for Radio Astrophysics of the Tata Institute of 
Fundamental Research. RK acknowledges support by the DST-INSPIRE Faculty Award.
We thank our colleagues for fruitful discussions on this system. This research has made use of 
the NASA/IPAC Extragalactic Database (NED) which is operated by the Jet Propulsion Laboratory, 
California Institute of Technology, under contract with the National Aeronautics and Space Administration. 
 The National Radio Astronomy Observatory is a facility of the National Science Foundation operated under cooperative 
 agreement by Associated Universities, Inc.



\bibliographystyle{mnras}
\bibliography{mybib_1}



\bsp	
\label{lastpage}
\end{document}